\documentclass[notitlepage,floatfix,twocolumn,aps,amsmath,amssymb,superscriptaddress,10pt]{revtex4-2}
\usepackage[pretty,uselistings,final]{revquantum}

\usepackage{comment}

\usepackage{xcolor}
\usepackage{soul}
\usepackage{graphicx}
\usepackage{adjustbox}
\usepackage{tablefootnote}
\usepackage{tabularx}
\usepackage{hyperref}
\usepackage{nameref}
\usepackage{datetime}

\definecolor{britishracinggreen}{rgb}{0.0, 0.26, 0.15}
\definecolor{forestgreen}{rgb}{0.13, 0.55, 0.13}
\definecolor{applegreen}{rgb}{0.55, 0.71, 0.0}
\definecolor{byzantine}{rgb}{0.74, 0.2, 0.64}
\definecolor{lapislazuli}{rgb}{0.15, 0.38, 0.61}
\definecolor{chocolate(web)}{rgb}{0.82, 0.41, 0.12}
\definecolor{blackpink}{rgb}{1,0.35,0.78}

\makeatletter
\newcommand{\thickhline}{%
    \noalign {\ifnum 0=`}\fi \hrule height 1pt
    \futurelet \reserved@a \@xhline
}
\newcolumntype{"}{@{\hskip\tabcolsep\vrule width 1pt\hskip\tabcolsep}}
\makeatother

\begin{document}

\title{
Single-emitter quantum key distribution over 175 km of fiber with optimised finite key rates}

\author{Christopher L. Morrison}
\email[Correspondence: ]{chrislmorrison93@outlook.com}
\affiliation{Institute of Photonics and Quantum Sciences, School of Engineering and Physical Sciences, Heriot-Watt University, Edinburgh EH14 4AS, UK}
\author{Roberto G. Pousa}
\affiliation{SUPA Department of Physics, University of Strathclyde, Glasgow G4 0NG, UK}
\author{Francesco Graffitti}
\affiliation{Institute of Photonics and Quantum Sciences, School of Engineering and Physical Sciences, Heriot-Watt University, Edinburgh EH14 4AS, UK}
\author{Zhe Xian Koong}
\affiliation{Institute of Photonics and Quantum Sciences, School of Engineering and Physical Sciences, Heriot-Watt University, Edinburgh EH14 4AS, UK}
\author{Peter Barrow}
\affiliation{Institute of Photonics and Quantum Sciences, School of Engineering and Physical Sciences, Heriot-Watt University, Edinburgh EH14 4AS, UK}
\author{Nick G. Stoltz}
\affiliation{Materials Department, University of California, Santa Barbara, California 93106, USA}
\author{Dirk Bouwmeester}
\affiliation{Huygens-Kamerlingh Onnes Laboratory, Leiden University, P.O. Box 9504, 2300 RA Leiden, Netherlands}
\affiliation{Department of Physics, University of California, Santa Barbara, California 93106, USA}
\author{John Jeffers}
\affiliation{SUPA Department of Physics, University of Strathclyde, Glasgow G4 0NG, UK}
\author{Daniel K. L. Oi}
\affiliation{SUPA Department of Physics, University of Strathclyde, Glasgow G4 0NG, UK}
\author{Brian D. Gerardot}
\affiliation{Institute of Photonics and Quantum Sciences, School of Engineering and Physical Sciences, Heriot-Watt University, Edinburgh EH14 4AS, UK}
\author{Alessandro Fedrizzi}
\affiliation{Institute of Photonics and Quantum Sciences, School of Engineering and Physical Sciences, Heriot-Watt University, Edinburgh EH14 4AS, UK}

\begin{abstract}
Quantum key distribution with solid-state single-photon emitters is gaining traction due to their rapidly improving performance and compatibility with future quantum network architectures. In this work, we perform fibre-based quantum key distribution with a quantum dot frequency-converted to telecom wavelength, achieving count rates of 1.6~MHz with $g^{\left(2\right)}\left(0\right) = 3.6 \%$. We demonstrate positive key rates up to 175~km in the asymptotic regime. We then show that the community standard analysis for non-decoy state QKD drastically overestimates the acquisition time required to generate secure finite keys. Our improved analysis using the multiplicative Chernoff bound reduces the required number of received signals by a factor of $10^8$ over existing work, with the finite key rate approaching the asymptotic limit at all achievable distances for acquisition times of one hour. Over a practical distance of 100~km we achieve a finite key rate of 13~kbps after one minute of integration time.
This result represents major progress towards the feasibility of long-distance single-emitter QKD networks.
\end{abstract}

\date{\today}

\maketitle
Future quantum networks will require bright low-noise sources of single photons to enable applications including secure communication and distributed quantum computing \cite{lu2021quantum}. There is a range of promising platforms for such a source, including quantum dots, molecules, quantum emitters in two-dimensional materials such as WSe$_2$ and hBN, and colour centres in wide band-gap materials such as diamond and SiC. Comparing these different possible platforms for single-photon emitters, quantum dots (QDs) have demonstrated the highest count rates with the lowest multiphoton emission probability~\cite{tomm_bright_2021,wang_towards_2019,Thomas21Bright}.

Fibre-based QKD requires single-photons at 1550~nm where loss in fibre is lowest. This can be realised with QDs in two ways, fabricating the QD to emit directly at 1550~nm or using quantum frequency-conversion to shift the wavelength of a QD which emits at shorter wavelengths to 1550~nm.
The best available QDs in all relevant metrics emit at shorter wavelengths \cite{tomm_bright_2021,wang_towards_2019,Thomas21Bright}, although recent improvements has been made with C-band emitters in terms of brightness and multiphoton noise but not coherence \cite{nawrath2022high}.
Quantum frequency-conversion has been shown to be a viable route to realise a bright, coherent telecom QD single-photon source with low multiphoton noise, leveraging the performance of shorter wavelength QDs~\cite{doi:10.1063/5.0045413, dalio2022pure, 600kmFreqConv}. 

\begin{figure*}[tb]
    \centering
    \includegraphics[width=1.85\columnwidth]{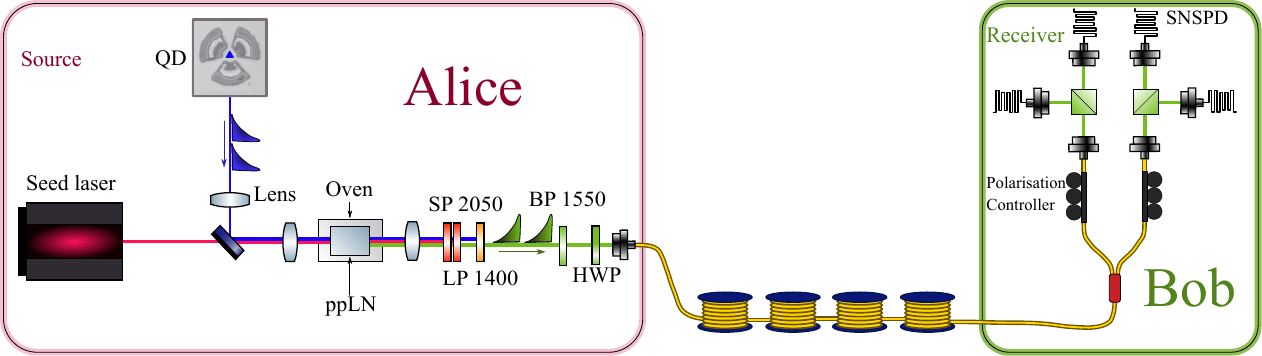}
    \caption{Experimental setup of Alice's source (pink outline) and Bob's passive BB84 receiver (green outline). The QD is excited at 160~MHz by temporally multiplexing an 80~MHz pulse train from  a Ti:Sapphire laser. The 940~nm single photons are converted to 1550~nm in a ppLN ridge waveguide designed to be single-mode at 1550~nm. The strong seed beam is removed with short-pass filters at 2050~nm (SP2050) before the telecom photons are isolated with a long pass filter at 1400~nm (LP1400) and a bandpass filter at 1550~nm (BP1550). The transmission channel consists of spools of fibre of various length which are joined using physical contact connectors for the different distances measured in Fig.~\ref{AsympKeyRate}. Bob's receiver passively chooses between X and Z basis measurements using a 50/50 fibre beam-splitter. Projections are made using polarising beam-splitter cubes and in-fibre polarisation controllers to align the measurement basis. }
    \label{exp_setup}
\end{figure*}

In this work we demonstrate Bennett-Brassard '84 (BB84) QKD~\cite{bb84} using a bright frequency-converted QD source over optical fibre. In the asymptotic case the source outperforms previous demonstrations of prepare and measure QKD  with single-photon emitters in terms of achievable key rate and maximum tolerable loss thanks to the brightness and low $g^{\left(2\right)}\left(0\right)$ of our source, see Table~\ref{otherWork}.  In the composable security framework, we use improved analytical bounds for the random sampling without replacement problem related to the phase error rate and the multiplicative Chernoff bound that has been proven to be a tighter finite key bound in other contexts~\cite{yin2020tight}. These bounds are used to calculate the fluctuations between expected and observed values.

The finite key treatment implemented in this work dramatically reduces the number of signals Bob must receive to approach the asymptotic case from $10^{15}$ to $10^{7}$ compared with previous single-photon source QKD analyses. Equivalently, the integration time required to approach the asymptotic case is reduced from $10^4$ years to just one hour. 

\section*{Results}
The source in this work consists of an InGaAs/GaAs quantum dot inside an oxide-apertured micropillar \cite{strauf_high_2007} emitting photons at 940~nm.
The QD is excited using a dark-field confocal microscope, single photons are collected in a cross-polarised scheme with $10^7$ suppression of the excitation laser. The QD is operated under pulsed quasi-resonant excitation using the third order cavity mode detuned by $440$~GHz from the QD emission. Femtosecond pulses from a Ti:Sapphire laser are stretched to 30~ps using a 4f Fourier pulse shaper and temporally multiplexed up to 160.7~MHz. We measure $\approx 5$~MHz count rate with a $g^{\left(2\right)}\left(0\right)=0.019(1)$ directly from the QD. 
The single photon emission is converted to 1550~nm in a difference frequency generation (DFG) process in a 48~mm periodically-poled lithium niobate (ppLN) waveguide pumped by a 2400~nm continuous-wave laser. The internal conversion efficiency of the DFG process is 57\%. Further details on the source can be found in \cite{doi:10.1063/5.0045413}. 

The four BB84 polarisation states $\left\{H,V,D,A\right\}$ are encoded using a motorised half-wave plate. Photons are then transmitted through the quantum channel consisting of SMF-28 fibre spools with an average propagation loss of $0.1904$~dB/km including connectors. The fibre is housed in an insulating box to reduce temperature fluctuations, which keeps the fibre-induced polarisation rotation stable over the typical acquisition time of 30 minutes per polarisation state.

\begin{figure*}[tb]
    \centering
    \includegraphics[width=1.95\columnwidth]{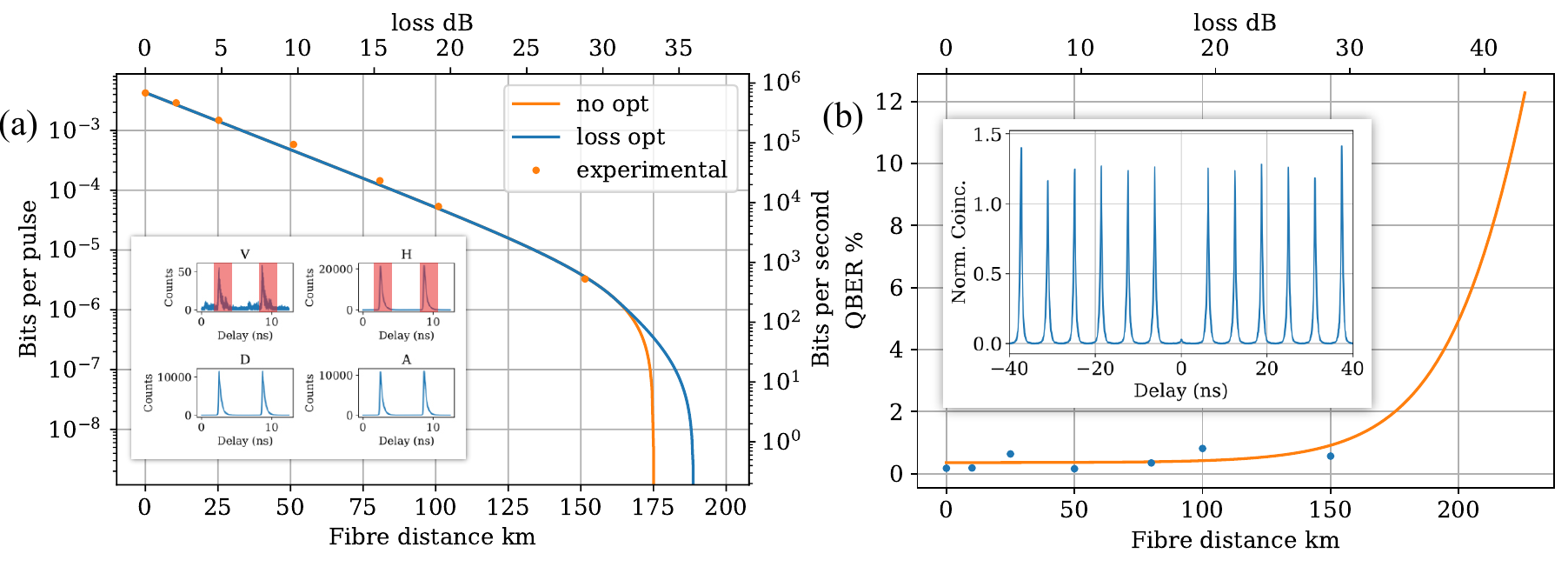}
    \caption{Asymptotic key rate and quantum bit error rate. \textbf{(a)} Experimental asymptotic key rate (orange dots) and the theoretical key rate based on the experimentally measured parameters with and without pre-attenuation of Alice's source. The pre-attenuation increases 2.6~dB the maximum tolerable loss. The inset shows a typical data set for Alice sending horizontally polarised photons over 80~km of fibre. Red boxes show the typical time gating used to optimise the key rate. 
    \textbf{(b)} Measured error rate as a function of fibre distance. The theory fit is based on Eq.~\ref{QBER_theory} with the experimental parameters listed in the main text. The deviations from the best fit are due to inconsistency in aligning the in-fibre polarisation controller. The QBER at the maximum tolerable loss of $\sim 35$ dB is $\sim 2\%$. Maximum tolerable loss is primarily limited by the photon-number noise $g^{\left(2\right)}\left(0\right) = 0.036(3)$, shown in the inset of \textbf{(b)}.}
    \label{AsympKeyRate}
\end{figure*}

The BB84 receiver consists of a 50/50 fibre beam-splitter followed by two polarising beam splitters and in-fibre polarisation controllers to project into the H/V and D/A basis respectively. Photons are detected with superconducting nanowire single-photon detectors (SNSPDs). The average transmittivity of the four arms of the receiver is 87\% including relative efficiency of each detector measured by comparing the count rate observed on each detector with a reference parametric down-conversion source. The SNSPDs are biased to have an average dark count rate of 11.5~Hz at the cost of 5-10\% of the peak efficiency. The detectors are time gated around the arrival time of the signal photons to reduce the effect of dark counts (see Fig. \ref{AsympKeyRate}), the average time gate across all distances is 3.19~ns. This gives a dark count probability per pulse of $p_{dc} = 1.47\times10^{-7}$. 

We send each of the BB84 states $\left\{H,V,D,A\right\}$ in turn and record at least $5\times 10^6$ detected events for each state for seven distances between 0-175~km. The probability that a given round registers in one of Bob's detectors, $p_{\text{click}}$, is estimated as the ratio of detected events to the number of clock pulses from the Ti:Sapphire which are recorded over the integration period. For convenience, we assume equal probabilities for both bases, i.e. $p_{dc} \equiv p_{dc}^{X} \equiv p_{dc}^{Z}$ and $p_{click} \equiv p_{click}^{X} \equiv p_{dc}^{Z}$. We measure a count rate of 1.6~MHz in Bob's receiver at zero distance. This gives a mean photon number of $\langle n\rangle=0.0142$ injected into the communication channel backing out the known receiver transmission, the relative efficiency on average due to the measured losses of each detector $\left(\approx87\%\right)$ and the estimated quantum efficiency of the detectors $\left(\approx75\%\right)$.

The quantum bit error rate (QBER) $e_{X/Z}$, in the X or Z basis is calculated by comparing the ratio of detected events for the state orthogonal to Alice's encoded state to the total number of detected events in that basis. By fitting the measured QBER to
\begin{equation}
    e_{X/Z} = \dfrac{p_{dc}/2 + p_{mis} \langle n\rangle T }{p_{dc} + \langle n\rangle T},
    \label{QBER_theory}
\end{equation}
the average polarisation misalignment $p_\text{mis}$ can be extracted, which typically is found to be $p_\text{mis} = 0.3\%$ \cite{yammamotoPRA}. $T$ represents the total optical efficiency from the quantum channel to Bob's detection apparatus. The dark count probability $p_{dc}$ and mean photon number $\langle n\rangle$ are held as fixed parameters.


With  $p_{\text{click}},\, e_{X/Z},\, \langle n\rangle$ and $g^{\left(2\right)}\left(0\right)$ experimentally characterised it is possible to calculate the asymptotic key rate (AKR) according to \cite{GLLP,Cai_2009}
\begin{equation}
    S = p_\text{sift} p_{\text{click}}\left[A \left(1-H\left(\dfrac{e_X}{A}\right)\right) - f_{EC}(e_Z)H(e_Z)\right],
\end{equation}
where $p_\text{sift}=p_X^2+(1-p_X)^2$ is the sifting ratio for the key generation bits assuming both bases are used, $p_X$ is the basis bias, $H\left(x\right)$ is the binary Shannon entropy and $f_{EC}\left(x\right) > 1$ is the error correction efficiency factor. 

For the experimental setup presented here $p_\text{sift}=\frac{1}{2}$ which allows for a comparison to previously published work (Table~\ref{otherWork}). For $f_{EC}\left(x\right)$ use values linearly interpolated between those reported in Ref.~\cite{PhysRevA.61.052304} (typically $f_{EC}=1.16$ for the range of error rates seen in the experiment). 
$A = \left(p_{\text{click}} - p_m\right)/p_{\text{click}}$ is the fraction of signals which are single-photon pulses and $p_m$ is the upper bound on the probability that Alice emits a multiphoton pulse taken to be $p_m \le g^{\left(2\right)}\left(0\right) \langle n\rangle^2 /2$~\cite{yammamotoPRA}. From the measured $\langle n\rangle = 0.0142$ and $g^{\left(2\right)}\left(0\right) = 0.036(3)$ (see Fig. \ref{AsympKeyRate}), we estimate $p_m\le 3.63\times10^{-6}$ without any additional pre-attenuation before the final collection fibre. The small increase in $g^{\left(2\right)}\left(0\right)$ compared to the emission directly from the QD is due to Raman scattering in the frequency-conversion process.

\begin{figure*}[tb!]
    \centering
    \includegraphics[width=1.95\columnwidth]{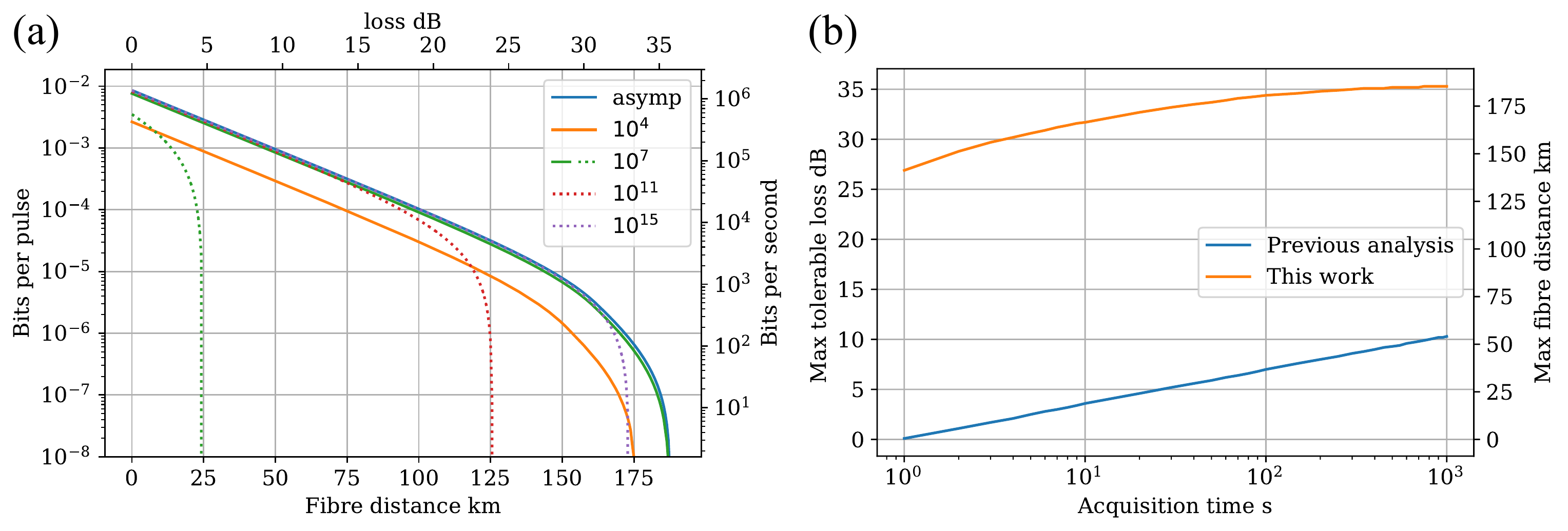}
    \caption{Comparison of finite key rate for the analysis presented in this work and previous analysis based on~\cite{Cai_2009}. The finite key rate for different block sizes is shown in  Fig \textbf{a}, the previous analysis is shown with dotted lines, the new analysis is shown with solid lines. For both versions of the analysis Alice's pre-attenuation and $p_\text{X}$ are optimised for each distance and integration time. The analysis presented in this work results in substantially better finite key rates using smaller block sizes. Fig \textbf{b} shows the maximum tolerable loss achievable as a function of the acquisition time. The new analysis substantially improves the distance over which a key can be generated particularly for short acquisition times.}
    \label{Finite key comparison}
\end{figure*}

The key rate at shorter distances is increased compared to previous works thanks to the high brightness and temporally multiplexed excitation presented in this work. The maximum tolerable loss is also increased due to the relatively high brightness and low noise compared to previous demonstrations with telecom wavelength QD sources. The current maximum range is limited by $p_{\text{click}} \rightarrow p_m$, at which point the fraction of signals received from single photon pulses goes to zero $A\rightarrow0$, and a secure key is no longer possible.  As the multiphoton emission and click probabilities are evaluated on a signal-by-signal basis, the multiplexed excitation does not improve the maximum distance over which a secure key can be extracted.

\begin{table}[htb]
\begin{adjustbox}{width=0.45\textwidth}
\begin{tabular}{l*{3}{l}}
\hline
\textrm{Reference}&
\textrm{AKR at 0~km (kbps)}&
\textrm{Maximum tolerable loss (dB)}\\
\hline
This work & 689 & 33.3\\
This work with active encoding$^*$ & 258 & 34.4\\
QD~\cite{takemoto_quantum_2015} & 4 & 23\\
QD~\cite{timmGao} & 2 & 23\\
QD~\cite{waks_quantum_2002}& 25 & 28 \\
Molecule~\cite{murtaza_efficient_2022}& 500 & 22  \\
2D Material~\cite{samaner2022free} & 0.24 & 21 \\
2D Material~\cite{gao2022atomically} & 150 & 23 \\
\hline
\end{tabular}
\end{adjustbox}
\caption{Comparison of other QKD demonstrations based on single-photon emitters. For the purposes of comparison, the asymptotic key rate has been calculated with $p_{sift}=\frac{1}{2}$ and with no additional source attenuation. Refs~\cite{takemoto_quantum_2015,samaner2022free} include active switching of the encoded state, all other demonstrations use static encoding. A thorough review of QKD with QDs can be found in~\cite{dotqkdreview}. $^*$Prediction based on 3~dB loss and 2\% polarisation encoding error typical with fibre-based electro-optic modulators. 
}
\label{otherWork}
\end{table}

In assessing the performance of a practical QKD system the finite key rate must be considered. For a finite block size defined as either the number of sent $N_S$ or received signals $N_R$, the total secure key length $\ell$ is,
\begin{align}
\label{eq:fkr}
    \ell = {} & \Bigl\lfloor \underbar{N}_{R,nmp}^X \left(1 - H\left(\bar{\phi}^X\right)\right)\nonumber\\
    & -\lambda_{EC} - 2\log_2 \frac{1}{2\varepsilon_{PA}} - \log_2 \frac{2}{\varepsilon_{cor}}\Bigr\rfloor,
\end{align}
where $\underbar{N}_{R,n}^X$ is the lower bound on the number of received signals in the key generation basis due to non-multiphoton source emissions (including vacuum and single-photon emissions), $\bar{\phi}^X$ is the upper bound of the phase error rate in the key generation basis, $\lambda_{EC}$ is the information leaked during error correction~\cite{tomamichel2017fundamental}, and the remaining terms are security  and correctness parameters derived using the methods in~\cite{bunandar2020numerical}. The key rate is then defined as $r=\frac{\ell}{N_S}$ and the fixed parameters used are shown in Table~\ref{Table_parameters}. The ratio of signals in the key generation basis to the parameter estimation basis, and the additional attenuation Alice adds to the source to reduce the multiphoton emission probability are all numerically optimised for each distance.

The improvement to the finite key rate can be viewed in two different ways: at long distances the block size required to produce the same key rate is massively reduced, Fig.~\ref{Finite key comparison}\textbf{a}; alternatively, for a fixed acquisition time we can tolerate more loss and achieve the same secure key rate, Fig.~\ref{Finite key comparison}\textbf{b}.
The improvement to the finite key rate is quantified by comparing the number of signals required to approach the asymptotic key rate. To approach the asymptotic key rate with the method of~\cite{Cai_2009}, the received block size has to be on the order of $10^{15}$, dotted purple curve Fig.~\ref{Finite key comparison}\textbf{a}. Our results indicates a factor $10^8$ improvement, the finite key rate curve reaches the asymptotic limit with Alice's pre-attenuation for just $10^7$ received signals. 
With respect to a fixed acquisition time, previous security analysis restricts the maximum distance over which a key can be exchanged after one second of acquisition time to less than 1~km, Fig.~\ref{Finite key comparison}\textbf{b}, whereas we calculate a maximum tolerable loss of 26.9~dB which is equivalent to over 140~km of fibre. For all acquisition times considered the analysis presented here can achieve the same key rate over an additional 25 dB of channel loss.

\begin{table}[t]
    \begin{adjustbox}{width=0.45\textwidth}
    \centering
    \begin{tabular}{l*{3}{l}}
        \hline
        \textbf{Description}\vspace{2pt} & \textbf{Parameter}\vspace{2pt} & \textbf{Value} \vspace{2pt} \\
        \hline
        Mean photon number & $\langle n\rangle$ & $0.0142$ \\
        Second-order correlation function & $g^{\left(2\right)}\left(0\right)$ & $0.036$ \\
        Source repetition rate & $R$ & $160.7$ MHz \\
        Misalignment probability & $p_{mis}$ & $0.003$ \\
        Dark count probability & $p_{dc}$ & $1.47\times10^{-7}$ \\
        Detector efficiency & $\eta_{det}$ & $0.6525$ \\
        Detector dead time & $\tau$ & $27.5$ ns \\
        Fibre loss & $l$ & $0.1904$ dB/km \\
        Parameter estimation failure probability & $\varepsilon_{PE}$ & $2 \times 10^{-10}/3$ \\
        Privacy amplification failure probability & $\varepsilon_{PA}$ & $10^{-10}/6$ \\
        Correctness failure probability & $\varepsilon_{cor}$ & $10^{-15}$ \\
        Error correction leakage & $\lambda_{EC}$ & Eq.~\ref{lambdaEC} \\
        \hline
    \end{tabular}
    \end{adjustbox}
    \caption{Baseline QKD system parameters.}
    \label{Table_parameters}
\end{table}

\section*{Discussion} 
We have demonstrated that fibre-based QKD with frequency-converted quantum dot is possible at high rates for distances and acquisition times relevant for metropolitan communication networks. The source performance exceeds other single-photon emitters suggested for use in QKD systems in terms of key rate and maximum tolerable loss. Combining state-of-the-art QD performance in brightness \cite{tomm_bright_2021} and multiphoton suppression \cite{veryLowg2}, with the frequency conversion demonstrated here into one device would allow for key rates comparable to decoy-state QKD with weak coherent pulses. Ultimately, surpassing weak coherent pulse implementations will require sources much closer to the ideal performance of unity collection efficiency with multiphoton emission probabilities approaching zero.

Regarding the key rate introduced with Eq.~\ref{eq:fkr}, a more up-to-date version for the terms of the security parameters and an additional fluctuation in the phase error rate due to the random sampling without replacement problem were introduced compared to previous studies. The considerable enhancement of the finite key rate is due to the improved bounds of the statistical fluctuations achieved using the Chernoff bound applied to the number of events versus bounding probabilities as in~\cite{cai2009erratum}. 

The deviations of the probabilities from the ideal estimate are magnified when expressed in the total number of events, e.g. number of errors and multiphoton emissions, although they might seem to be relatively small. In particular, the Chernoff bound on events provides tighter estimates on the maximum number of multiphoton emissions increasing the single photon yield at longer distances and consequently the key rate.

\section*{Methods}

\subsection*{Click and Error Probability Estimation}
\label{sec:clicks}

In this section, we describe the modelling of click probabilities and error rates to later simulate the detections and error events. First, the click probability in each basis is,
\begin{equation}
    p_{c}^{X,Z} = c_{dt}\sum_{n=0}^{\infty} p_n \left[ 1 - (1 - p_{dc}^{X,Z}) \left(1 - \eta_{ch} \eta_{det}^{X,Z} \eta_{att} \right)^n\right],
\end{equation}
where $p_n$ is the probability that a pulse emitted by the source contains $n$ photons, $\eta_{det}^{X,Z}$ is the detector detection efficiency and $p_{dc}^{X,Z}$ is the average dark count probability of the two detectors associated with each basis. For simplicity we assume that all detectors have the same efficiencies and dark count rates. If they differ, then the security analysis should be adapted to avoid any loopholes introduced by detector efficiency mismatch~\cite{bunandar2020numerical}. We add a pre-attenuation factor $\eta_{att}$~\cite{yammamotoPRA} which can be inserted between the source and the Eve-controlled channel to reduce multiphoton leakage in the high loss regime. The channel transmittance is given by $\eta_{ch} = 10^{-l/10}$ where $l$ is the channel loss in  $dB$. A correction factor $c_{DT}$ is  added to account for the dead time of the detectors. For a dead time $\tau$ and repetition rate $R$ this correction is of the form
\begin{equation}
    c_{dt} = \frac{1}{1 + R \tau p_{c}^{X,Z}}.
\end{equation}
The error probability is then given by
\begin{equation}
\begin{split}
    p_{e}^{X,Z} &= c_{dt} \left\{ p_0 \, p_{dc}^{X,Z} \right. \\
    &+\left. \sum_{n=1}^{\infty} p_n \left[ 1 - (1 - p_{dc}^{X,Z}) (1 - \eta_{ch} \eta_{det}^{X,Z} \eta_{att})^n\right]  p_{mis} \right\},
    \end{split}
\end{equation}
where $p_{mis}$ is the probability of error due to the misalignment of the set-up.

For modelling purposes, we will assume that the multiphoton contribution is dominated by the 2-photon component, hence consider a source distribution of the form $\{p_n\}=\{p_0, p_1, p_2\}$ with emission probabilities of vacuum $p_0$, single photons $p_1$ and two photon states $p_2$. Given mean values for photon number $\langle n\rangle$ and $g^{\left(2\right)}\left(0\right)$,
\begin{align}
        p_2 &= \frac{g^{(2)} \langle n\rangle^2}{2},\quad
        p_1 = \langle n\rangle - 2 p_2,\quad
        p_0 = 1 - p_2 - p_1.
\end{align}
Note that the security of the key rate analysis is not compromised by such an assumed form of the photon number distribution as the distribution that only has non-zero $\{p_0,p_1,p_2\}$ saturates the bound of~\cite{yammamotoPRA}, 
\begin{equation}
\label{pmYamamoto}
    p_m \le \frac{g^{\left(2\right)}\left(0\right) \langle n\rangle^2}{2},
\end{equation}
and any other distribution consistent with $\langle n\rangle$ and $g^{\left(2\right)}\left(0\right)$ will have a lower $p_m$.

\subsection*{Finite Key Length based on Chernoff Bounds}
\label{sec:chernoff}

In this section, we follow the method and notation as described in~\cite{yin2020tight} though suitably adapted for the single-photon source case. 

After basis sifting, the number of events where both Alice and Bob chose the Z and X bases are $N_R^X= N_S p_X^2 p_c^X$ and $N_R^Z= N_S p_Z^2 p_c^Z$ respectively. Here, we adopt the convention that the $Z$ basis is used for parameter estimation and the $X$ basis is used to generate the key. The legitimate parties publically compare all the $Z$ basis results to determine the number of $Z$ errors $m_Z = N_S p_Z^2 p_e^Z$ which is then used to estimate the phase error rate $\phi^X$ in the $X$ basis. The $X$ basis results are never directly revealed.

The expected number of received signals that result from non-multiphoton emissions by Alice (lumping together the vacuum and single photon yields) is given by $N_{R,nmp}^{X,Z} = N_R^{X,Z}-N_{S,mp}^{X,Z*}$ where $N_{S,mp}^{X,Z*}$ is the expected number (we use ${}^*$ to denote the mean) of multiphoton emissions from Alice in the $X,Z$ basis respectively. Here, we assume that all multiphoton pulses are detected by Bob (Eve introducing a lossless channel in this case) and that the remaining detected pulses come from the non-multiphoton fraction (if $N_R^{X,Z}>N_{S,mp}^{X,Z}$). As we do not directly observe the actual number of $Z$ multiphoton emissions, the actual number $N_{S,mp}^{X,Z}$ can deviate from ${N_{S,mp}^{X,Z*}}$ due to statistical fluctuations, and we need to upper bound the tail probability with error $\varepsilon_{PE}$. The upper Chernoff bound (denoted by the overbar) for a sum of binary variables $x=\sum x_j$ with $ x_j\in\{0,1\}$ is given by
\begin{equation}
    \bar{x}=(1+\delta^U) x^*,
\end{equation}
where $\delta^U=\frac{\beta+\sqrt{8\beta x^* + \beta^2}}{2x^*},$ and $\beta=-\log_e (\varepsilon_{PE})$. This can be applied to derive an upper bound to the actual number of multiphoton emissions $\bar{N}_{S,mp}^{X,Z}$, hence lower bound the number of received signals from non-multiphoton emission events, $\underbar{N}_{R,nm}^{X,Z}$ in each basis,
\begin{equation}
    \underbar{N}_{R,nmp}^{X,Z}=N_R^{X,Z}-\bar{N}_{S,m}^{X,Z}.
\end{equation}

The phase error rate $\phi^X$ now needs to be upper bounded based on the observed number of errors in the $Z$ basis $m_Z$. We conservatively assume that all $Z$ basis errors occur on the received non-multiphoton fraction, hence we have an estimate of the phase error rate,
\begin{equation}
    \phi^X=\frac{m_Z}{\underbar{N}_{R,nmp}^{Z}}.
\end{equation}
However, this estimate is the result of $N_R^Z$ samples in the $Z$ basis but we need to upper bound the phase error rate in the unannounced $N_R^X$ samples in the $X$ (key generating) basis. For this random sampling without replacement problem and a tail bound error $\varepsilon$, the upper bound of the unobserved value $\chi$ can be estimated from the observed value $\lambda$ by
\begin{equation}
    \chi=\lambda+\gamma^U \left( n, k, \lambda, \varepsilon' \right),\\
    \end{equation}
where
\begin{eqnarray}
    \gamma^U\left( n, k, \lambda, \varepsilon' \right)&=&\frac{1}{2+2\frac{A^2 G}{(n+k)^2}}\left\{\frac{(1-2\lambda)AG}{n+k}\right.\\ \nonumber
    &&+\left. \sqrt{\frac{A^2 G^2}{(n+k)^2}+4\lambda(1-\lambda)G}\right\},\\
    A&=&\text{max}\{n,k\},\\
    G&=&\frac{n+k}{nk}\log_e \frac{n+k}{2\pi nk\lambda(1-\lambda)\varepsilon'^{\,2}},
\end{eqnarray}
under the assumption that $0<\lambda<\chi<0.5$ which is true for typical QKD scenarios. This now allows us to calculate an upper bound, 
\begin{equation}
    \bar{\phi}^X = \phi^X + \gamma^U \left( N_R^X, N_R^Z, \phi^X, \frac{\varepsilon_{sec}}{6} \right).
\end{equation}
The secrecy of the protocol is $\varepsilon_{sec} \geq \varepsilon_{PA} + \varepsilon_{PE} + \varepsilon_{EC}$ where: $\varepsilon_{PA} = \varepsilon'$ is the privacy amplification failure probability; $\varepsilon_{PE} = 2 n_{PE} \varepsilon'$ is the parameter estimation failure probability where $n_{PE} = 2$ is the number of constraints as quantified in post-processing; $\varepsilon_{EC} = \varepsilon'$ is the error correction failure probability. Thus, the secrecy comes from setting each failure probability to a common value $\varepsilon'$, i.e. $\varepsilon_{sec} = 6 \varepsilon'$. Moreover, the QKD protocol is $\varepsilon_{qkd}$-secure if it is $\varepsilon_{cor}$-correct and $\varepsilon_{sec}$-secret with $\varepsilon_{qkd} \geq \varepsilon_{cor} + \varepsilon_{sec}$. We set $\varepsilon_{cor} = 10^{-15}$ and $\varepsilon_{sec} = 10^{-10}$.


This leads to the length of the secure key fraction,
\begin{align}
    \ell = {} & \Bigl\lfloor \underbar{N}_{R,nmp}^X \left(1-H\left(\bar{\phi}^X\right)\right)\nonumber \\
    {} & -\lambda_{EC} - 2\log_2 \frac{1}{2\varepsilon_{PA}} - \log_2 \frac{2}{\varepsilon_{cor}}\Bigr\rfloor,
\end{align}
where $\lambda_{EC}$ is the known leakage of information during error correction. The key rate is then defined as $r=\frac{\ell}{N_S}$.

\subsection*{Security bounds and secure key rate}
\label{sec:bounds}

The security analysis follows that of~\cite{bunandar2020numerical} using min-entropy and the failure probabilities that appear in Table~\ref{Table_parameters} therein. We use uncertainty relations for bounding Bob's raw key obtained from Alice's raw key and conditioned on Eve's information. Let us first consider Eve's information $E$ and Alice's raw key $X_A$, that is generated by choosing a random sample from $n_X$, after the error correction and verification steps. The question is how much Eve information can extract from $X_A$ that is completely unknown to her. The probability of guessing $X_A$ given $E$ is defined as the classical min-entropy,
\begin{equation}
\label{Hmin}
    H_{min}\left( X_A | E \right) = \log_2 p_{guess}\left( X_A | E \right),
\end{equation}
where $p_{guess}\left( X_A | E \right)$ represents the probability of correctly guessing $X_A$ applying an optimal extraction strategy having access to $E$. The optimal strategy means to guess the value $x$ of $X$ with the highest conditional probability $p_{X|E=e}(x)$ for each value $e$ of $E$. For this process, let us assume that a part $X_B$ of $X_A$ with length $\ell$, that is uniform conditioned on the information $E$, can be extracted by Bob. In other words, there is a function $f_s$ that maps $X_A$ to Bob's raw key $X_B = f_s(X_A)$ considering the quantum state between Alice and Eve $\rho_{X_A E}$ is fixed. It has been shown that the probability of guessing $X_B$ is $p_{guess}\left( X_B | E \right) = 2^{-\ell}$ and using eq. (\ref{Hmin}) we obtain,
\begin{equation}
    H_{min}\left( X_B | E \right) = \ell,
\end{equation}
where $l$ is the secure key length. Furthermore, because $X_B$ comes from mapping $X_A$, the probability of correctly guessing $X_B$ has to be greater than the probability of guessing $X_A$. Therefore, these min-entropies can be expressed as the following inequality
\begin{equation}
    \label{Hmin1}
    H_{min}\left( X_B | E \right) \leq H_{min}\left( X_A | E \right) \Rightarrow \ell \leq H_{min}\left( X_A | E \right).
\end{equation}
To extend this to the general case of almost uniform randomness, the smooth min-entropy $H_{min}^{\varepsilon}\left( X_A | E \right)$ needs to be introduced. This is set as the maximum value of $H_{min}\left( X_A | E \right)$. For privacy amplification, we consider that Alice and Bob apply a two-universal hash function. The \textit{Leftover Hashing Lemma}~\cite{tomamichel2011leftover} gives us an exact equation for the inequality of eq.~(\ref{Hmin1}) using the smooth min-entropy to relate the already mentioned Eve's information $E$ and Alice's raw key $X_A$
\begin{equation}
    \ell = H_{min}^{\varepsilon}\left( X_A | E \right) - 2 \, \log_2 \frac{1}{2 \varepsilon_{PA}}
\end{equation}
for the maximum number of extractable bits $l$ that are $\varepsilon_{PA}$-close to uniform, conditioned on $E$.

We consider leakage $\lambda_{EC}$ during error correction as well as additional bits for verification. Thus, the information that remains in Eve's system $E'$ after error correction is related by,
\begin{equation}
    H_{min}^{\varepsilon}\left( X_A | E \right) \geq H_{min}^{\varepsilon}\left( X_A | E' \right) - \lambda_{EC} - \log_2 \frac{2}{\varepsilon_{cor}}.
\end{equation}
The leakage in one-way protocols is lower bounded as~\cite{tomamichel2017fundamental},
\begin{align}
\label{lambdaEC}
\begin{split}
    \lambda_{EC} & \geq  n_X H(e_X) \\
    & + \left[ n_X \left( 1 - e_X \right)- F^{-1} \left( \varepsilon_{cor};\, n_X,\, 1-e_X \right)\right] \log_2 \frac{1-e_X}{e_X}  \\
    & - \frac{1}{2}\log_2 n_X - \log_2 \frac{1}{\varepsilon_{cor}},
\end{split}
\end{align}
where $H(x)$ is the binary Shannon entropy, and $F^{-1} \left( \varepsilon_{cor}; \,n_X, \,1-e_X \right)$ is the inverse of the cumulative distribution of the binomial distribution. Achievable rates by practical codes may not achieve this bound for large blocks so we choose the greater estimate of leakage given either by the above or $f_{EC}=1.16$~\cite{PhysRevA.61.052304}.

We use an uncertainty relation for smooth min-entropy to establish a bound between the remaining information that Eve has, $E'$, and Alice's raw key, $X_A$. This reflects that the better Bob can estimate Alice's raw key in the $Z$ basis, the worse Eve can guess Alice's raw key in the $X$ basis, formally expressed as,
\begin{equation}
    H_{min}^{\varepsilon}\left( X_A | E' \right) \geq q \, s_{X,nm} - H_{max}^{\varepsilon}\left( Z_A | Z_B \right),
\end{equation}
limited to the non-multiphoton events in the key generation basis X. Here, $q$ quantifies the efficiency of Bob's measurements, in our case $q=1$ as Bob uses orthogonal bases. $H_{max}^{\varepsilon}\left( Z_A | Z_B \right)$ is the smooth max-entropy of $Z_B$ conditioned on $Z_A$. If $Z_B$ and $Z_A$ are highly correlated, we can deduce that $H_{max}^{\varepsilon}\left( Z_A | Z_B \right)$ is small and thus, as the following bound shows (see Lemma 3 of~\cite{bunandar2020numerical}), the observed number of errors is small,
\begin{equation}
    H_{max}^{\varepsilon}\left( Z_A | Z_B \right) \leq s_{X,nm} H(\phi_X),
\end{equation}
where $\phi_X$ is the X-basis phase error rate of non-multiphoton events. Finally, the bound for the min-entropy is,
\begin{equation}
    H_{min}^{\varepsilon}\left( X_A | E' \right) \geq s_{X,nm} \left[ 1 - H(\phi_X) \right].
\end{equation}

\subsection*{Protocol optimisation}
\label{sec:optimisation}

To maximise the rate and tolerable loss whilst maintaining security, we consider optimisations of the basis bias and signal pre-attenuation that can provide some improvement over standard protocol values, i.e. equal basis choice and no-attenuation.

The Efficient BB84 protocol simplifies standard BB84 by utilising one basis for key generation and the other basis for parameter estimation of the phase error rate, without compromising security~\cite{lo2005efficient}. In this paper, we adopt the convention that the $X$ basis is used for the key with the $Z$ basis used for phase error rate estimation. Alice and Bob randomly and independently choose their basis for each signal with bias $p_X$ and $p_Z=(1-p_X)$. The sifting ratio is $1-2p_X(1-p_X)>\frac{1}{2}$ for unequal bias, higher than the sifting ratio $\frac{1}{2}$ for $p_X=\frac{1}{2}$ as in standard BB84. Additionally, this simplification also reduces the number of parameters to be estimated, hence improving finite-statistical bounds and the reduction in key length due to composable security parameters~\cite{bunandar2020numerical,sidhu2021key,brougham2021medium,brougham2022modelling,sidhu2022finite}. The value of $p_X$ can be optimised to balance the amount of raw key bits (proportional to $p_X^2$) and parameter estimation signals (proportional to $(1-p_X)^2$). In the asymptotic limit, $p_X\rightarrow 1$, hence the sifting ratio also approaches unity.

At long distances and high losses, the key rate is limited by the multi-photon emission probability. When the upper bound on the number of multiphoton emission events exceeds the number of detections, then Eve must be assumed to have full information about Alice and Bob's string, hence there can be no secure key. Waks et al.~\cite{yammamotoPRA} proposed the addition of linear attenuation (characterised by transmission factor $\eta_{att}$) of the signals prior to injection into the quantum channel controlled by Eve. The bound on the multiphoton components is reduced by a factor of $\eta_{att}^2$ whilst the average photon number is only reduced by $\eta_{att}$. At high losses and with low dark count rates, the reduction in detection probability (and increase in QBER) may be offset by the greater fraction of Bob's received events being the result of non-multiphoton emissions by Alice, potentially leading to increased key rate and extending the non-zero key rate region to longer ranges.

\bibliography{reference.bib}

\section*{Acknowledgements}
D.K.L.O. is supported by the EPSRC Researcher in Residence programme at the Satellite Applications Catapult (EP/T517288/1). R.G.P. acknowledges support from the EPSRC Research Excellence Award (REA) Studentship. D.K.L.O. and R.G.P. are supported by the EPSRC International Network in Space Quantum Technologies (EP/W027011/1). J.J. is supported by QuantIC, the EPSRC Quantum Technology Hub in Quantum Imaging (EP/T00097X/1).
A.F., D.K.L.O., and R.G.P. are supported by the EPSRC Quantum Technology Hub in Quantum Communication (EP/T001011/1). B.D.G. is supported by a Wolfson Merit Award from the Royal Society, a Chair in Emerging Technology from the Royal Academy of Engineering, and the ERC (grant no. 725920).
F.G. and Z.X.K. acknowledge studentship funding from EPSRC under Grant No. EP/L015110/1.

\section*{Author contributions}

C.L.M., F.G. and Z.X.K. performed the measurements and collected the experimental data. P.B. assisted with data analysis. N.G.S. and D.B. fabricated the quantum dot sample. R.G.P., J.J., and D.K.L.O. derived the finite key bounds and performed the finite key optimisation. B.D.G. and A.F. conceived and supervised the experiment. All authors contributed to writing the manuscript.

\end{document}